\documentclass{aa}
\usepackage{graphicx}
\begin{document}
\title{ GIRAFFE multiple integral field units at VLT: a unique tool to recover
velocity fields of distant galaxies\thanks{Based on ESO GTO programs No 71.A-0322(A) and 72.A-0169(A) } }


\author{H. Flores\inst{1}
          \and
          M. Puech\inst{1}
          \and
          F. Hammer\inst{1}
          \and
          O. Garrido\inst{1}
          \and
          O. Hernandez\inst{2,3}
          }

\offprints{hector.flores@obspm.fr}
\authorrunning{H. Flores et al.}
\titlerunning{Velocity field of distant galaxies}

\institute{GEPI, Observatoire de Paris Meudon, 92190 Meudon, France
         \and
         Lab. d'Astrophysique de Marseille, France
        \and
Lab. d'Astrophysique Exp\'erimentale 
et Obs. du mont M\'egantic, U. de Montr\'eal, Qu\'e., Canada H3C 3J7
        }
\date{---}

\abstract{ The GIRAFFE spectrograph is unique in providing the
      integral field spectroscopy of fifteen distant galaxies at the
      same time. It has been successfully implemented at the second
      VLT unit within the FLAMES facility. We present GIRAFFE
      observations acquired during the Guaranteed Time
      Observation of the Paris Observatory, using total exposure times
      ranging from 6 to 12 hours. The reduced 3D cube of each galaxy
      has been deconvolved using our new package DisGal3D. This
      software has been written using the only assumption that UV
      light traces the emission line regions. The comparison between
      GIRAFFE spectra and HST imagery allows us to recover details on
      velocity fields as small as $0.3-0.4$ arcsec. It has been
      successfully tested using Fabry Perot observations of nearby
      galaxies purposely redshifted to large
      distances. We present here preliminary results for three distant
      galaxies at $0.45< z < 0.65$, whose velocity fields have
      been derived with exquisite spectral ($R=10000$) and spatial
      resolutions. Observed velocity fields range from disturbed fields
      expected in major merger events to those of regular spiral with
      minor perturbations.  For the latter, one could accurately
      derive the dynamical major axis and the maximal rotational
      velocity. We conclude that dynamical properties of a large
      number of distant galaxies can be routinely derived at VLT.
      This opens a new avenue towards the understanding of the galaxy
      formation and evolution during the last 8 Gyr.

       }

\maketitle
%

\section{Introduction}
Studies of galaxies at intermediate redshift ($0.4 < z < 1.2$) have
revealed large changes of galaxy properties during the last 8 Gyr,
which follow the strong declines of the cosmic star formation density
(Lilly et al., 1996; Flores et al., 1999) and of the merging rate (Le
F\`evre et al., 2000). Indeed, galaxies at intermediate redshift have
complex morphologies and colors differents from those of the Hubble
sequence and they show metal abundances lower than those of present
day galaxies (Hammer et al., 2004). Major contributors for this
evolution have been identified to be luminous IR galaxies (LIRGs) and
luminous compact galaxies (Flores et al., 1999; Hammer et al., 2001).
Besides the numerous studies of their photometric and chemical
properties, very little is known about the dynamical properties of
galaxies beyond $z=0.1$.

The Tully-Fisher relation is hard to reproduce in simulations
(Steinmetz \& Navarro 1999), and it is of prime importance to study
its evolution until $z=1$. Significant changes of the slope of the
Tully-Fisher relation are expected in the distant Universe ($z\sim 1$,
Ferreras and Silk, 2001) and this would be the best observable to
explore the star formation process in disk galaxies. The TF relation
at high redshift has been investigated by several studies using slit
spectroscopy (Simard \& Pritchet 1998; Bershady et al., 1999; Vogt et
al., 2000; Barden et al., 2003; Boehm et al., 2003). Most of them have
revealed a brightening of the rest frame B-band, but its magnitude
(from $0.2$ to $1.1$ mag) as well as the constraints on the TF slope
(see Fig. 10 in Ferreras \& Silk) are far from being predictive up to
now. Kinematics studies using 3D spectroscopy appear to be a
pre-requisite to sample the whole Velocity Field (hereafter VF) of
individual galaxies to distinguish between interacting and
non-interacting galaxies (Mendes de Oliveira et al., 2003), and to
limit uncertainties related to the major axis determination. Such
effects have been already tested for nearby spirals by comparing
Fabry-Perot (hereafter FP) observations with long-slit spectroscopies.
The latter can easily provide under or over estimates of the maximum
velocity by factors reaching 50\% (Amram et al., 1995). Could slit
spectroscopy be appropriate for distant galaxies which are actively
forming stars (up to rates larger than $100$ M$_\odot /$yr) and where
the frequency of interactions is very common ?  It seems that our
present knowledge of the dynamics of distant galaxies is very poor,
maybe comparable to that provided from long slit spectroscopy of local
galaxies in the beginning of the last century (Wolf, 1914).  During
the last 20 years, 2D velocity fields obtained from scanning FP
interferometers (and more recently also from integral field and side
by side long slit spectroscopy) have proved to be powerful kinematic
tools to investigate the properties of nearby galaxies (Veilleux et
al., 2001; Kosugi et al., 1995; Swinbank et al., 2003; Mendes de
Oliveira et al., 2003; Garrido et al., 2004; Ostlin et al., 2001).

To observe VF of distant galaxies, the integral mode IFU of
FLAMES/GIRAFFE at VLT seems particularly well suited. Compared to
available integral field instruments (e.g. GEMINI/GMOS, Swinbank et
al.,  2003), the FLAMES/GIRAFFE instrument and its IFU mode provide 3D
spectroscopy of fifteen distant galaxies at the same time on a 20
arcmin FoV, thus this instrument optimizes the long exposure time
needed to observe faint objects. Moreover, with IFUs, spectra are
directly observed within each pixel and no further analysis is needed
in contrary of FP data.
 
In this paper, we present the preliminary analysis of three (among 50)
distant galaxies up to $z=0.7$ which have been observed with GIRAFFE/IFU
(section 2).  In section 3 we present our new package {\bf DisGal3D}
which has been purposely developed to derive galaxy VF using the
combination of GIRAFFE/IFU spectroscopy and HST/F606W imagery.  A
major purpose of this paper is to derive the accuracy of our method 
determining the maximal velocity as well as  distinguishing disturbed
VF from those, more regular, typical of spiral galaxies. In section 4
we present our first results on distant galaxies.

\begin{table*}
\caption{ Observational and deduced parameters of the three distant galaxies.}
\begin{center}
{\scriptsize
\begin{tabular}{cccccccccccccc}\hline
CFRS    & z     & I     & M$_B$  &  f$_{\lambda}^a$ & $(f_{\lambda}/A)^a$ & T$_{\rm int}$& S/N$^b$  & Morph$^c$ &  VF$^d$ & D.P.A$^e$ & O.P.A.$^f$ & i$^f$ &  V$^e_{\rm max}$ \\\hline
03.0508 & 0.464 & 21.9 & $-$20.01 & 31.0 & 11.2 & 6hr & 446 &  Sp   &  Sp     & 119$\pm 5$& 143$\pm 2$& 39$\pm 3$& 70 \\
03.1309 & 0.617 & 20.6 & $-$21.76 & 14.9 & 8.0 & 12hr & 182 &Merger &  Merger & N.A. & N.A. & N.A. & N.A.  \\
03.9003 & 0.619 & 20.8 & $-$21.51 & 21.3 & 1.5 & 12hr & 310 & Sp/Ir& Warp Sp & 296$\pm 5$& 326$\pm 2$& 52$\pm 3$& 190 \\\hline
\end{tabular}
}
\end{center}
\begin{list}{}{}
\item[$^{\mathrm{a}}$] [OII] integrated  from VLT/FORS spectroscopy and [OII] per arcsec$^2$. Fluxes in units of $\times10^{-16}$ ergs/s/cm$^2$/{\rm \AA} and $\times10^{-16}$ ergs/s/cm$^2$/{\rm \AA}/arcsec$^2$   using HST images to measure the surface of object.
\item[$^{\mathrm{b}}$] Total S/N from reconstructed spectrum using the 20 IFU channel spectra. 
\item[$^{\mathrm{c}}$] From HST images (Brinchman et al., 1998; Zheng et al., 2004).
\item[$^{\mathrm{d}}$] Dynamical morphology from reconstructed velocity field.
\item[$^{\mathrm{e}}$] Principal axis and maximum velocity deducted using ADHOCw package (http://www-obs.cnrs-mrs.fr/adhoc/adhoc.html).
\item[$^{\mathrm{f}}$] Optical principal axis and inclination deducted from HST I-band images.
\end{list}
\end{table*}


\begin{figure*}[ht!]
\centering
\caption{ (JPEG figure) Example of performed DisGal3D tests with nearby galaxies
(here NGC 2403, observed with  FanTOmM $R=27000$, Hernandez et al., 2004 et Gach et al., 2003). (a): H$\alpha$ map used during the deconvolution
process (see text). (b): VF obtained by FP observation at the
mont Megantic Observatory ($1.6$ arcsec/pix). (c): same VF obtained with the
galaxy redshifted to ${z\sim 0.22}$ (0.1 arcsec/pix corresponding to the
HST/WFPC2 sampling). (d): deconvolved VF obtained by DisGal3D after
convolution of the redshifted galaxy by a $0.6$ arcsec seeing (see
text). Superimposed isovelocities are those from (c).}
\label{Fig1}
\end{figure*}

\section{Observations}
As part of Guaranteed Time Observation programs of the Paris
Observatory (P.I: F.  Hammer) more than 50 distant galaxies have been
observed, using the ESO VLT/FLAMES facility, IFU mode ($3''\times 2''$
array of 20 square $0.52''$ width microlenses) with setups L04
($R=0.55$ {\rm \AA} -- $30$ km/s) and L05 ($R=0.45$ {\rm \AA} --
$22$ km/s), and integration times from 4 to 12 hours (ESO runs
No 71.A-0322(A) and 72.A-0169(A)). Observational seeing during
1hr $\times$ n exposures was ranging from $0.4$ to $0.8$ arcsec.  Data
reduction has been done using the dedicated software BLDRS developed
at the Geneva Observatory (http://girbldrs.sourceforge.netweb). Sky
subtraction has been made using standard IRAF and written purpose IDL
tools. The three galaxies presented here have been selected from their
morphological properties, ranging from an apparently well formed disk
(03.0508) to an extreme case of merging (03.1309). Table 1 summarizes
observational strategy and deduced parameters of the three distant
galaxies presented in this paper.

\section{Analysis}
VF of galaxies have been reconstructed using a dedicated IDL package
named DisGal3D. DisGal3D includes a guided microscanning algorithm and
a standard deconvolution method to reconstruct VF using 3D
observations. It will be detailed in a forthcoming paper (Puech et al.,
2004 in preparation, hereafter P04). Our method has been validated
using ten FP cubes of nearby galaxies observed at the mont Megantic
(Hernandez et al., 2004 in prep) and Haute Provence Observatories
(Garrido et al., 2004). Galaxies have been redshifted in order to
simulate GIRAFFE IFU spatial observational conditions
but no specific spectral treatment has been done. Several simulated
GIRAFFE data cubes at different seeings (from $0.6$ to $1$) have been
produced.

Our deconvolution method proceeds in a two main steps algorithm. The
first one consists of interpolating the GIRAFFE data cube through a
guided microscanning thanks to HST/F606W images (Chemin et al., 2003).
Here is the single assumption of our method, that is, the emission
line regions are traced by the UV light. Indeed both UV and [OII] are
directly or indirectly associated to hot star emissions and their
luminosities correlates well (Swinbank et al., 2003; Hammer \& Flores,
1998).  Notice that [OII] line fall in the F606 filter at $z=0.4-0.8$.
At this stage, only the spatial sampling has been increased and the
spatial resolution remains still unchanged. The second step consists
of a standard Maximum Entropy deconvolution method of each spatial
slice of the interpolated cube. Spatial resolution is then improved
down to approximatively one/half a GIRAFFE pixel ($0.52''$). Thus, no
kinematical assumption has been made, since our method is not
model-dependent. We then do background subtraction, line
symmetrisation and central wavelength estimation by a barycenter
method (Garrido et al., 2002). Our software include classical box
smoothing or a filtering process based on wavelets decomposition
(P04).

\begin{figure*}[ht!]
   \centering
\caption{(JPEG figure) Up: positions of the $3\times 2$ arcsec$^2$ IFU bundle on sky, superimposed to
HST I-band images. Down: $3 \times 2$ arcsec isovelocities of observed galaxies derived
with DisGal3D. Each velocity field has been previously cleaned
spatially by a sky level thresholding and spectrally by a S/N line
thresholding. All features identified as possible
extrapolation/deconvolution artifacts are removed before computing
isovelocites within ADHOCw. Photometry in background is HST V-band,
used during the deconvolution process (see text). The dynamical axis
is much better defined than the optical axis, which is color dependent
and may be affected by many irregularities.}
\label{Fig2}
\end{figure*}

Simulated deconvolved GIRAFFE cubes have been then compared to
original FP observations (Figure 1). Our analysis
demonstrates that GIRAFFE VF are equivalent to smoothed FP
observations. This method allows us to discriminate between disturbed
and spiral like galaxies.  Figure 1 shows one of the nearby galaxy
used to test our software.  A preliminary analysis of residuals
between Figure 1c and Figure 1d indicates that structures with sizes
$> 0.3-0.5$ arcsec are recovered (P04). A preliminary study of deducted rotation curves shows
that $\Delta$V should be recovered within a 10\% error (P04).

\section{Results}   
To derive the VF of each galaxy we have used the well resolved
[OII][$3726.2,3728.9$]{\rm \AA} doublet emission lines. Figure 2
displays the three optical I-band images (upper panel) superimposed to
the IFU bundle and the respective reconstructed isovelocities (lower
panel) superimposed to high resolution HST images in rest frame
UV-band. CFRS03.0508 presents a regular VF with isovelocity lines
which suggest a possible warp of the E-N side of the disk, probably
due to the companion at the same redshift, $60$ kpc off and opposite
to it. CFRS03.9003 was classified by Brinchmann et al.  (1998) as a
spiral galaxy. However after analysing its colormap, Zheng et
al. (2004) classify it as an irregular. It is forming stars at a very
high rate, $\sim$ 75 M$_\odot/$yr (Flores et al., 2004). At $z=0.619$,
CFRS03.9003 present a symetric VF with a peculiar pattern of the
isovelocity lines. For the two above galaxies we have been able to
derive their rotation curves using the ADHOCw package
(http://www-obs.cnrs-mrs.fr/adhoc/adhoc.html, see Table 1 and Figure
3). To derive a rotation curve from a VF, one needs to correct from
projection effects on the plane of the sky. The simplest way to do it
is to first estimate the kinematical parameters $i$ (inclination angle
of the galaxy), $PA$ (Position Angle of the dynamical major axis),
${(X_{C}, Y_{C})}$ (dynamical center) and ${V_{\rm sys}}$ (systemic
velocity). Afterwards, these parameters are visually optimized within
ADHOCw until the rotation curve reaches a sufficient degree of
symmetry (at least in the rise of the curve if asymmetries occur in
the VF). In our case, estimates are based on photometric PA, center
and inclination (outer isophot determined by Sextractor). The most
spectacular galaxy is a giant merger, CFRS031309 at $z=0.617$ (Flores
et al, 2004). The VF is extremely irregular and somewhat chaotic as
revealed by sharp variations of the velocity along this ``chain''
galaxy. These can be taken as evidences for strong interactions,
meaning that this system is evolving rapidly as proved also by its
very high star formation rate ($\sim$ 200 M$_\odot/$yr, Flores et al.,
2004).

\section{Conclusion}
From a preliminary analysis of three isolated field galaxies observed
with the FLAMES facility (GIRAFFE/IFU mode), we show its efficiency in
producing VF of distant galaxies. Our main conclusions are:\newline
-- 3D spectroscopy with GIRAFFE IFUs is able to distinguish disturbed
VF from those of regular spirals;\newline
-- maximal velocity of distant regular spirals can be estimated within
$\Delta V=10$\%, an accuracy which can be only obtained with 3D
spectroscopy;\newline
-- with its 15 deployable IFUs, VLT/GIRAFFE is the best tool to
establish a robust Tully Fischer relation up to $z=1.2$, which is
independent of galaxy interactions or of crude assumptions on the VF
(major axis, inclination, barycenter).

 Further tests must be done about the capability to distinguish finer
 dynamical details such as minor merger, bars and warps. This is
 related to the spatial accuracy of our observations and of our
 deconvolution techniques. We will also investigate whether (heavy)
 extinction can affect our results. Establishing the VF of several
 hundreds of distant galaxies is now within the capabilities of the VLT and this is a
 new tool to investigate how the Hubble sequence was formed. A
 preliminary public release of DisGal3D will be soon available.

\begin{figure}[ht!]
   \centering
\caption{ (JPEG figure) Preliminary rotational velocity curves (corrected from
projection effects) of CFRS03.0508 (up) and CFRS03.9003 (down)
deducted using ADHOCw (preceding sides are in blue/empty
circles). Both reveal a perturbed side due to possible disturbations
in the VF. Each point/error bar represents the average/scatter
obtained on a 1 pixel crown ($\sim 0.1$ arcsec) spreading over given
angular sector centered on $PA$ (respectively 40 and 36 degrees). In
the case of CFRS03.0508, the decreasing side is likely due to an
interaction with the nearby companion.}
\label{Fig3}
\end{figure}

\begin{acknowledgements}
  We are especially indebted to P. Amram and C. Carignan who provide
  us with several datacubes of velocity fields of nearby galaxies,
  including those from the GHASP project. We thank I. Fuentes-Carrera
  and C. Balkowski for useful discussions and advices. We wish to
  thank the excellent work of the GIRAFFE team at Paris-Meudon
  Obseervatory as well as at ESO and Geneva Observatory (the FLAMES
  consortium).  H.F. and M.P. wish to thank the ESO Paranal staff for
  their reception and very useful advices.
\end{acknowledgements}

\end{document}